\documentclass[12pt,prd,aps,superscriptaddress,preprintnumbers,amsmath,amssymb,nofootinbib]{revtex4-1}
\pdfoutput=1
\usepackage{amsmath,graphics,epsfig}
\usepackage{color,hyperref}
\usepackage{datetime}
\usepackage{booktabs}
\usepackage{color}

\begin{document}

\title{Scalar Dark Matter: Real vs Complex }
\author{Hongyan Wu}
\affiliation{Department of Physics, Chongqing University, Chongqing 401331, P. R. China}
\author{ Sibo Zheng}
\affiliation{Department of Physics, Chongqing University, Chongqing 401331, P. R. China}
\date{October 20,  2016}

\begin{abstract}
We update the parameter spaces for both a real and complex scalar dark matter via the Higgs portal.
In the light of constraints arising from the LUX 2016 data, the latest Higgs invisible decay and the gamma ray spectrum,
the dark matter resonant mass region is further restricted to a narrow window between $54.9-62.3$ GeV in both cases,
and its large mass region is excluded until $834$ GeV and $3473$ GeV for the real and complex scalar, respectively.
\end{abstract}

\maketitle

\section{Introduction}
Recently, a few new limits on Dark Matter (DM) direct detection were released.
For example,
in comparison with the LUX 2015 data \cite{LUX2015}
there is about a factor of $\sim 4$ improvement on the DM-nucleon scattering cross section in the latest LUX 2016 data \cite{LUX2016}. 
This improvement may impose significant effects on simplified DM models with DM mass around the electroweak scale.
DM via the Higgs portal is a type of such simplest examples.
Usually, the number of model parameters in such DM sector is quite small,
making it very sensitive to the LUX experiments. 
The aim of this paper is to explore the implications of nowadays experimental limits to the Higgs-portal DM models.

In this scenario, Standard Model (SM) Higgs mediates interactions between DM and SM sector, with DM directly coupled to the SM Higgs via Yukawa interaction.
They are only two model parameters in the effective Lagrangian - the DM mass and the ``effective" Yukawa coupling constant.
Once the constraint from DM relic abundance is imposed,
the number of independent model parameters is reduced to one.
Since previous experimental data has excluded a fermionic DM (except for the pseudo-scalar coupling) \cite{1402.6287}, 
we focus on either a real or complex scalar DM,
on which early studies can be found in 
\cite{real1, real2, real3, real4,
real5, real6, real7, real8, real9, real10, real11, real12,
real13, real14, real15, real16, real17}
and \cite{complex, 
1005.3328, 1202.1316,1205.6416,1301.2599,1405.3530,1512.05355} ,
respectively \footnote{For a review on these models, see, e.g, \cite{1306.4710}.}.

In the light of the direct detection arising from the LUX 2016 data, 
the indirect detection arising from the Higgs invisible decay at LHC \cite{1606.02266},
and the indirect detection arising from gamma ray spectrum induced by DM annihilation at Fermi-LAT \cite{1506.00013, 1503.02641} and HESS \cite{1301.1173},
the DM mass regions are updated as follows. 
$(i)$ For the real scalar DM the resonant mass and large mass region is modified to $54.9 \leq m_{s} \leq 62.3$ GeV and $m_{s}\geq 834$ GeV, respectively.
$(ii)$ For the complex scalar DM the resonant mass is similar to the real scalar DM and the large mass region is modified to $m_{s}\geq 3473$ GeV.
See Table.\ref{summary} for details.
Although we will not discuss $N$ copies of real scalar with $N\geq 3$, 
it can be inferred that the large mass region for these choices is excluded, 
and the resonant mass region will be further suppressed as $N$ increases.

The plan of this paper is organized as follows.
Sec. II is devoted to address the notation in the real and complex DM models.
In Sec. III we discuss the direct detection at the LUX experiments, where the update will be noted.
In Sec. IV we discuss the indirect detections at the LHC, Fermi-LAT and HESS.
If available, we will compare the differences between the real and complex scalar DM.
Finally, we conclude in Sec.V.

\section{Scalar Dark Matter}
The Lagrangian for the dark sector in the DM model is given by,
\begin{eqnarray}{\label{Lag1}}
\mathcal{L}_{\text{dark}}=\frac{1}{2}\partial_{\mu} s\partial^{\mu} s^{*}- \frac{\mu^{2}}{2}\mid s\mid^{2}- \frac{\lambda}{2}\mid s\mid^{4} -\frac{\kappa}{2}\mid s\mid^{2}\mid H\mid^{2},
\end{eqnarray}
where $\mu $ is the DM bare mass,
and the last term denotes the interaction between DM sector and SM Higgs,
with $\kappa$ referring to the Yukawa coupling constant.
In order to keep the DM stable and eliminate harmful operators,
one simply imposes the following $Z_2$ parity on the full Lagrangian, 
\begin{eqnarray}{\label{parity}}
s \rightarrow -s,~~~~ \text{SM~particles} \rightarrow \text{SM~particles}.
\end{eqnarray}
Below the electroweak scale, Eq.(\ref{Lag1}) is rewritten as,
\begin{eqnarray}{\label{Lag2}}
\mathcal{L}_{\text{dark}}=\partial_{\mu} s\partial^{\mu} s^{*}- \frac{m_{s}^{2}}{2} \mid s \mid^{2}-\frac{\lambda}{2}\mid s\mid ^{4} -\frac{\kappa\upsilon}{2}\mid s\mid^{2}h - \frac{\kappa}{4}\mid s\mid^{2}h^{2}.
\end{eqnarray}
Here $m^{2}_{s}=\mu ^{2}+\kappa\upsilon^{2}/2$ is the square of DM mass,
and $H=(\upsilon +h)/\sqrt{2}$,  with the electroweak  scale $\upsilon\simeq 246$ GeV.

Dark matter $s$ can be composed of a single or multiple real scalar components. 
In this paper, we focus on the following two models,
\begin{eqnarray}{\label{def}}
s=\left\{
\begin{array}{lcl}
s_{1},~~~~~~~~~~~\text{Model~A}\\
s_{1}+is_{2},~~~~\text{Model~B}
\end{array} \right. 
\end{eqnarray}
which corresponds to a real and complex scalar, respectively. 
Note that the masses for $s_{1,2}$ in the model B are degenerate.
In some situation beyond Eq.(\ref{Lag1}), it may include a small mass mixing term,
which directly leads to non-degenerate masses. 
Mixing effects will be neglected in the following discussion.

\section{Direct Detection}
Let us firstly discuss the direct detection on the model A and B in the light of the latest LUX data.
Previous discussions can be found in Ref. \cite{1609.03551} and Ref.\cite{1609.09079} for the model A and B, respectively.
Consider that the self-interaction of DM is decoupled from the signals at both particle colliders and DM direct detection experiments,
only the DM mass $m_{s}$ and the Yukawa coupling constant $\kappa$ are sensitive to the LUX experiments.
The spin-independent DM-nucleon scattering cross section depends on these parameters as,
\begin{eqnarray}{\label{crosssection}}
\sigma_{\text{SI}}=c(i)\times\frac{\kappa^{2}f^{2}_{N}\bar{\mu}^{2}m^{2}_{N}}{4\pi m^{4}_{h}m_{s}^{2}},
\end{eqnarray}
where $m_{N}$ is the nucleon mass,
$\bar{\mu}=m_{N}m_{s}/(m_{N}+m_{s})$ is the DM-nucleon reduced mass,
and $f_{N}$ is the hadron matrix element.
Note that  the factor $c(A)=1$ and $c(B)=2$ for the model A and B, respectively.

\begin{figure}
\includegraphics[width=0.45\textwidth]{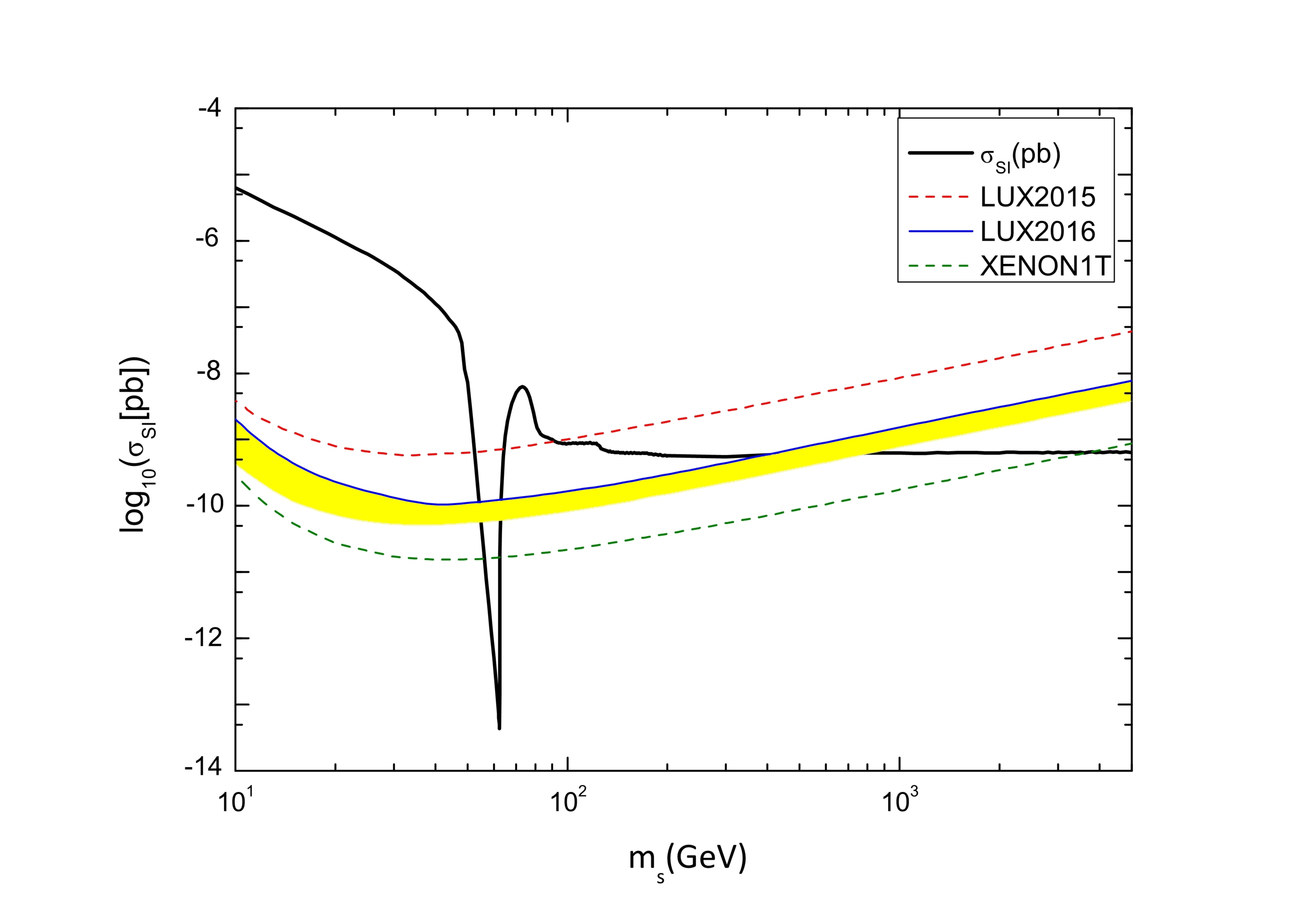}
\vspace{-0.3cm}
\includegraphics[width=0.45\textwidth]{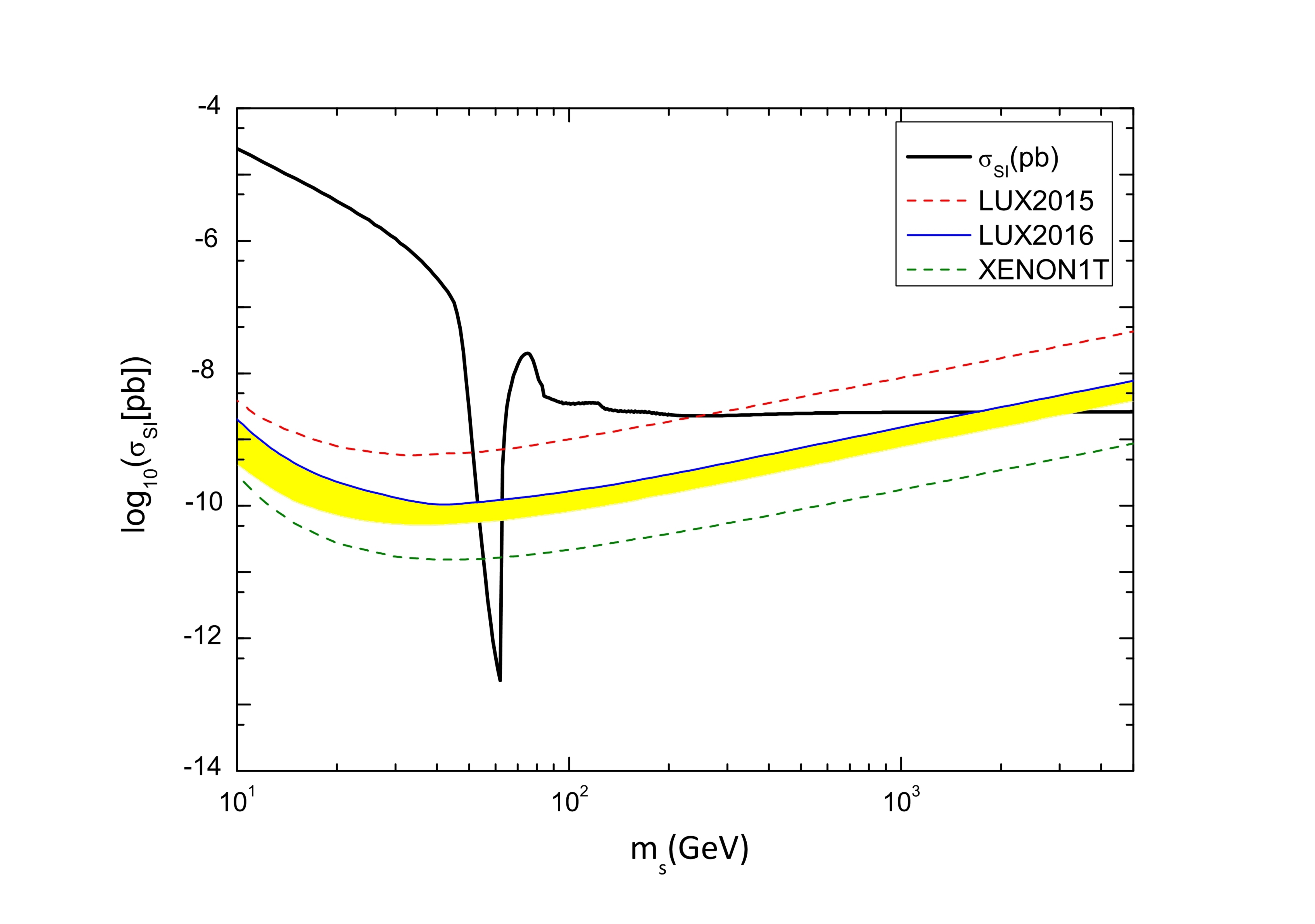}
\vspace{-0.3cm}
 \caption{DM-nucleon scattering cross section $\sigma_{SI}$ as the function of DM mass in the model A ($\mathbf{left}$) and model B ($\mathbf{right}$),
 respectively.
The dependence of $\sigma_{SI}$ on Yukawa coupling constant $\kappa$ is eliminated by the constraint from DM relic abundance as shown in Table.\ref{input}.
Curves of LUX 2015 (red), LUX 2016 (blue) and Xenon1T (green) \cite{Xenon1T} are shown simultaneously,
and the yellow band refers to $2\sigma$ deviation to their central values that were chosen by authors in Ref.\cite{1609.03551, 1609.09079}.}
 \label{lux}
\end{figure}

In fig.\ref{lux} we show the plots of $\sigma_{\text{SI}}$ as the function of DM mass by virtue of the code MicrOMEGAs \cite{1407.6129},
where the dependence of $\sigma_{SI}$ on Yukawa coupling constant $\kappa$ in Eq.(\ref{crosssection}) is eliminated by the constraint from DM relic abundance.
Input values for parameters in Eq.(\ref{crosssection}) are shown in Table \ref{input}.
In this figure, both the curves of LUX 2015 (red) and 2016 (blue) data are shown simultaneously.
Note that for the purpose of exclusion we have used their experimental values corresponding to $2\sigma$ deviations to their central values,
which is different from Refs. \cite{1609.03551, 1609.09079}.
Consequently, the ability of exclusion is expected to be stronger in our discussions.

For the model A, we find that the large mass region is obviously uplifted to $834$ GeV (LUX 2016) from $185$ GeV (LUX 2015).
In contrast, there is only about ~$\sim 1$ GeV deviation in the resonant mass region,  
which is modified from $53.5-63.5$ GeV (LUX 2015) \cite{1601.06232} to $54.9-62.8$ GeV (LUX 2016).
Meanwhile, for the model B, we find that the large mass region is obviously uplifted to $3473$ GeV (LUX 2016) from $247$ GeV (LUX 2015),
and the deviation in the resonant mass region is small similar to model A. 

\begin{table}
\begin{center}
\begin{tabular}{c} \hline
$\Omega_{\text{DM}}h^{2}=0.1199\pm 0.0027$ \cite{1303.5076}  \\
$f_{N}\simeq0.3$ \cite{1306.4710} \\
$m_{h}=125$ GeV \cite{1207.7214,1207.7235}  \\
\hline
\end{tabular}
\caption{DM relic abundance and input values for parameters in Eq.(\ref{crosssection}).}
\label{input}
\end{center}
\end{table}

In comparison with the model A, the DM mass lower bound in the large mass region is larger in the model B.
There are two reasons for this result.
At first, the contribution to DM relic abundance in the model B is roughly doubled given the same DM mass, in compared with model A. 
In order to reproduce the required DM relic density as shown in Table \ref{input}, 
the Yukawa coupling $\kappa$ should be multiplied by $\sqrt{2}$, 
as inferred from that $\Omega_{\text{DM}}h^{2}$ is proportional to $1/\kappa^{2}$.
We have verified this by the numerical calculations in terms of MicrOMEGAs.
Second, the factor $c(B)$ is two times of $c(A)$.
Therefore, the DM-nucleon scattering cross section in the model B is roughly $\sim 4$ times of that in the model A in the large mass region.
However, in the resonant mass region where $\sigma_{\text{SI}}$ is small, the enhancement effect is not so obvious as in the large mass region.

Other direct detections on either the resonant mass region or the large mass region in fig.\ref{lux} are rather insufficient.
The discovery of collider signatures requires extremely large integrated luminosity at least of order 
$\mathcal{O}(10)$ $ab^{-1}$ at the 14-TeV LHC \cite{1601.06232}.
The discovery of astrophysical signatures requires the DM scattering cross section relative to the DM mass,  
$\sigma/M$ at least of order $10^{-7}\text{cm}^{2}/g$ \cite{1510.06165}, 
in contrast to present limits typically of order $0.1\text{cm}^{2}/g$. 
Some cosmological considerations may impose interesting constraints on these models.
See, e.g., Ref. \cite{1509.01765}.

\section{Indirect Detection}
In this section we discuss indirect detections arising from Higgs invisible decay at the LHC as well as 
the gamma ray spectrum at the Fermi-LAT and HESS.
In contrast to the large mass region, these experiments may impose strong constraints on the resonant mass region.

\begin{figure}
\includegraphics[width=0.45\textwidth]{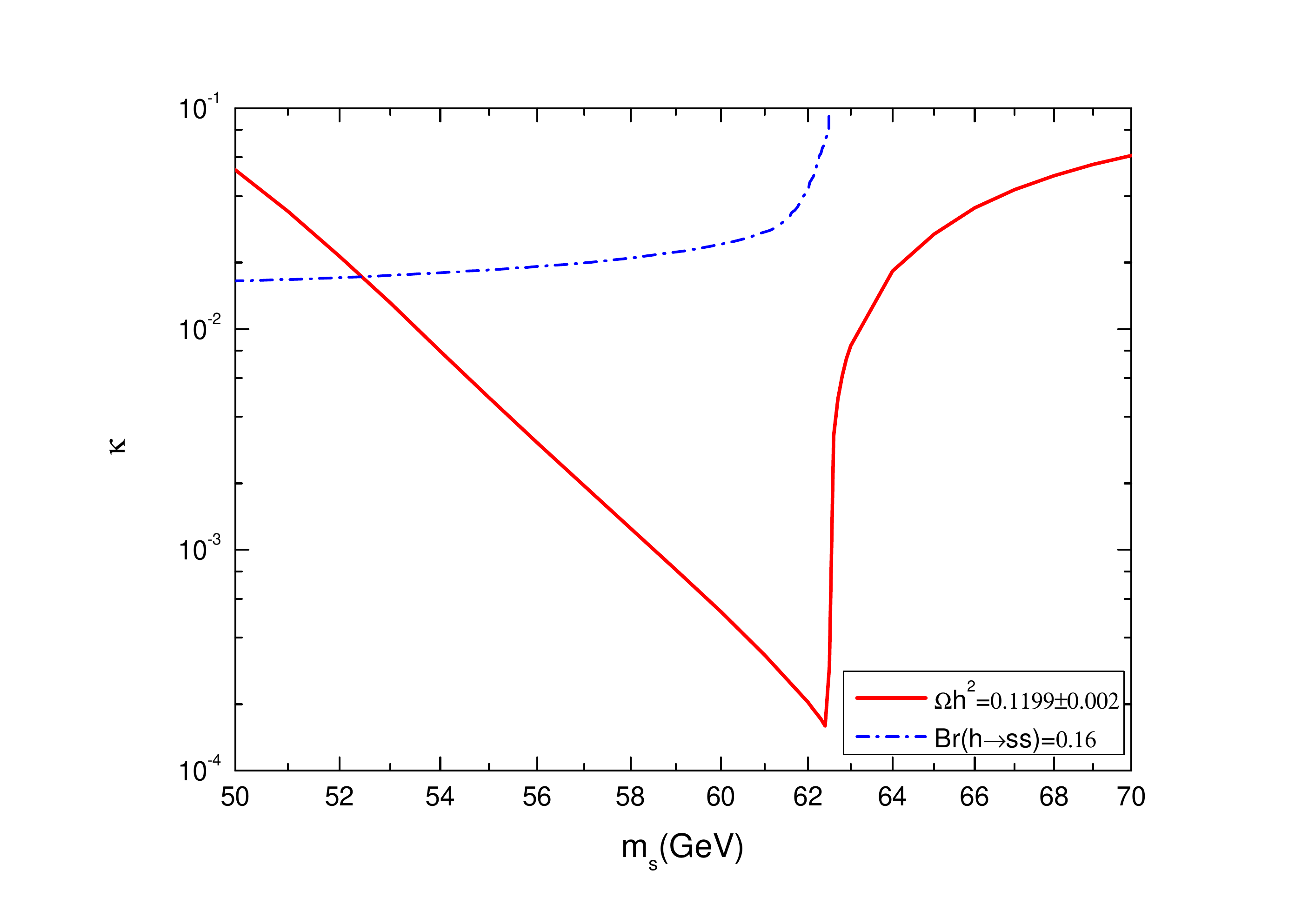}
\vspace{-0.3cm}
\includegraphics[width=0.45\textwidth]{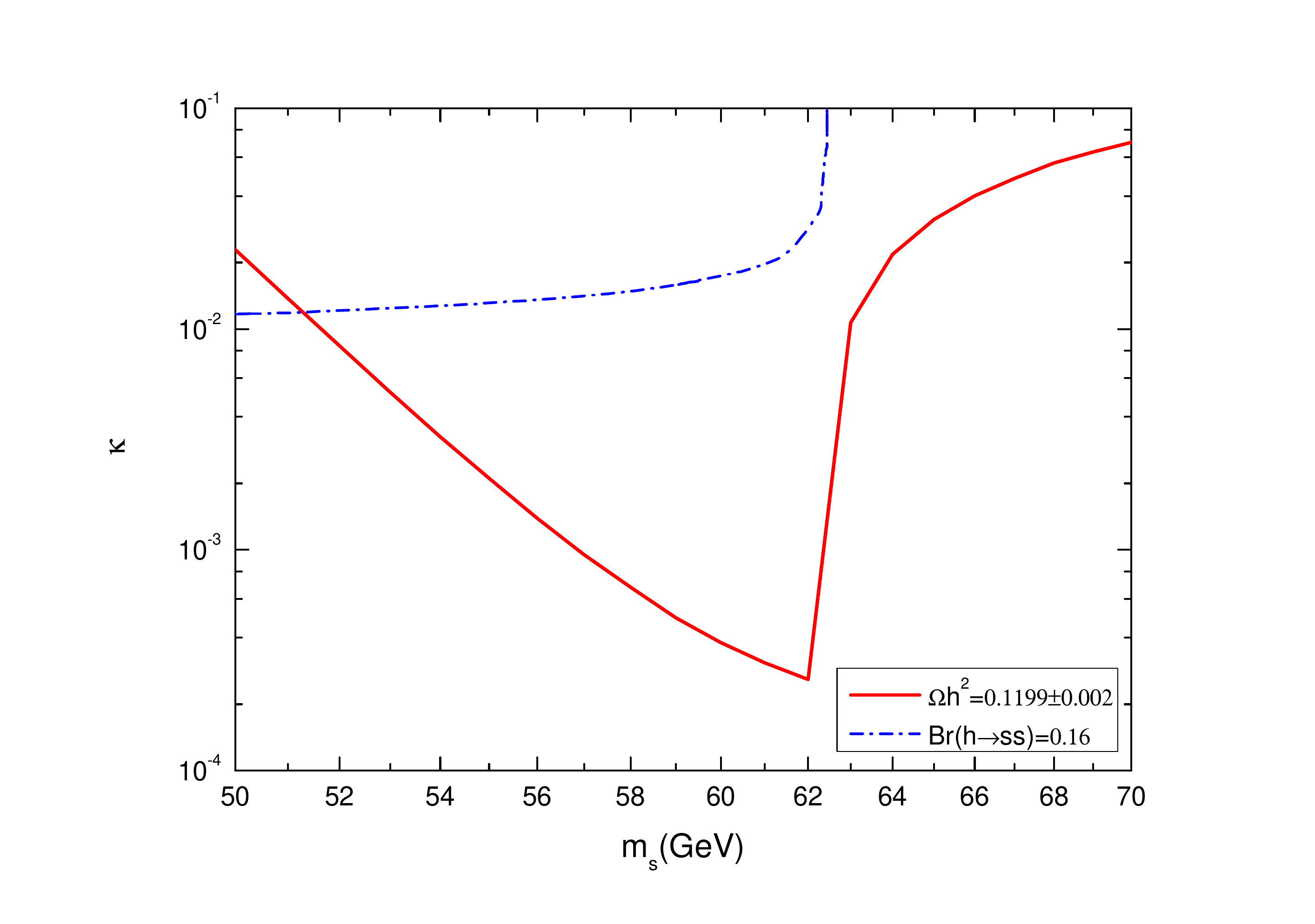}
\vspace{-0.3cm}
 \caption{Constraint from SM Higgs invisible decay at the LHC in the model A ($\mathbf{left}$) and model B ($\mathbf{right}$), respectively. 
 The red curve in each plot corresponds to the DM relic abundance.}
 \label{decay}
\end{figure}

\subsection{Higgs invisible decay}
When the DM mass is smaller than $m_{h}/2$, 
the measured Higgs invisible decay at the LHC imposes strong constraint on the decay width $\Gamma(h\rightarrow ss)$.
For details on the calculation of $\Gamma(h\rightarrow ss)$, see, e.g., \cite{1509.01765}.
The latest LHC result has been updated as \cite{1606.02266},
\begin{eqnarray}{\label{invisible}}
\Gamma(h\rightarrow ss) \leq 0.16 ~\Gamma_{h},
\end{eqnarray}
where the SM Higgs decay width $\Gamma_{h}\simeq 4.15$ MeV.
In fig.\ref{decay} we show this constraint on the DM mass,
which indicates that for the model A the DM mass lower bound in the resonant region has been modified to $52.3$ GeV from $51.8$ GeV \cite{1601.06232},
and for the model B the value of DM mass lower bound in the same region is about $51.5$ GeV.

\subsection{Gamma ray}
Now we discuss the indirect detection arising from gamma ray spectrum at Fermi-LAT and HESS.
These experiments impose upper bounds on the magnitudes of thermal averaged DM annihilation cross sections times velocity $\upsilon_{\text{rel}}$.
Consider that the maximal value of these cross sections in our models corresponds to the DM mass near $m_{h}/2$, 
the resonant mass region is mostly sensitive to these experiment limits.
In the following discussion we will use the gamma ray limits on $<\sigma_{\gamma\gamma} \upsilon_{\text{rel}}>$,
for which earlier analysis in the model A can be found in \cite{1412.1105, 1509.04282, 1603.08228, 1604.04589}.

\begin{figure}
\includegraphics[width=0.45\textwidth]{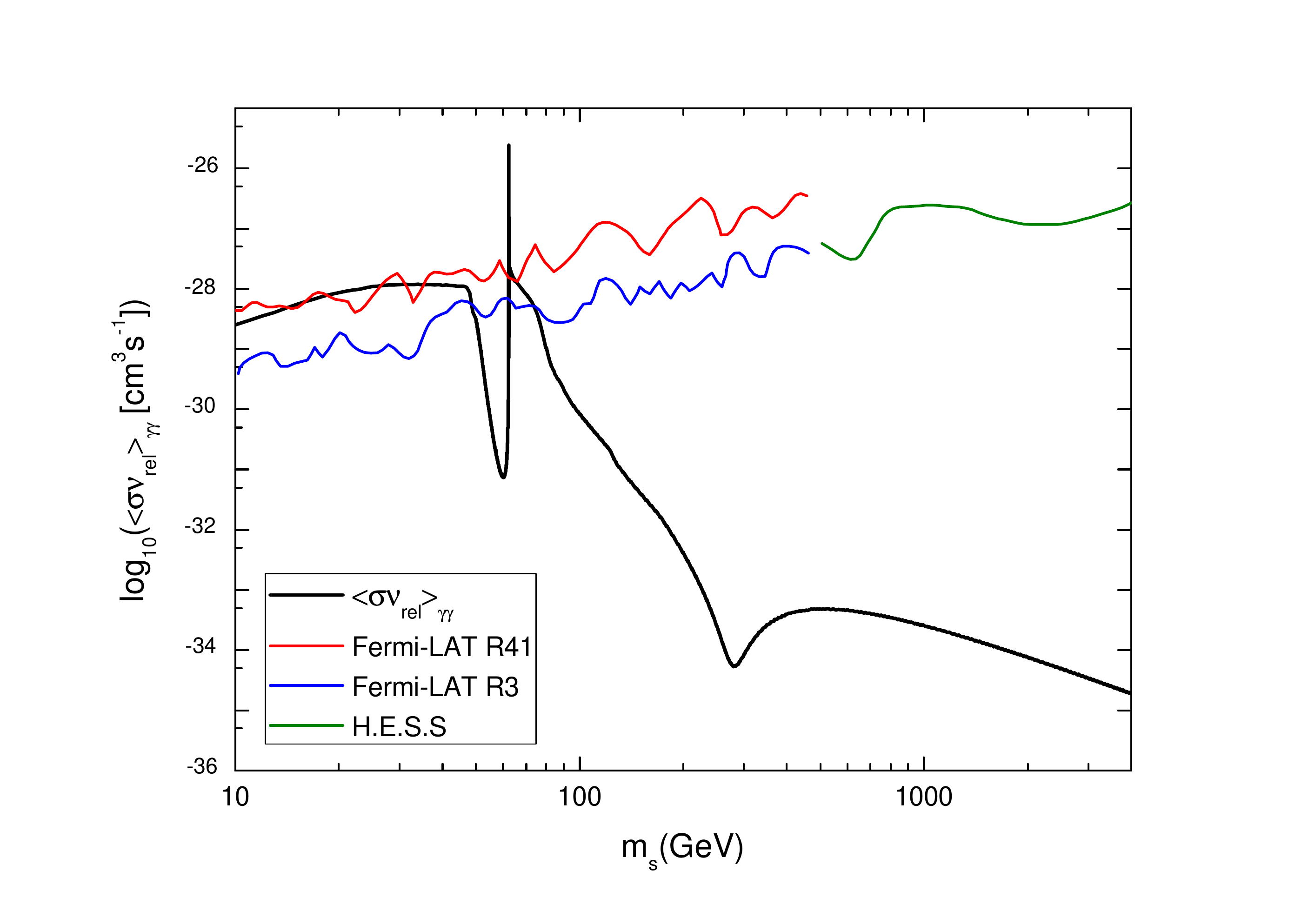}
\vspace{-0.3cm}
\includegraphics[width=0.45\textwidth]{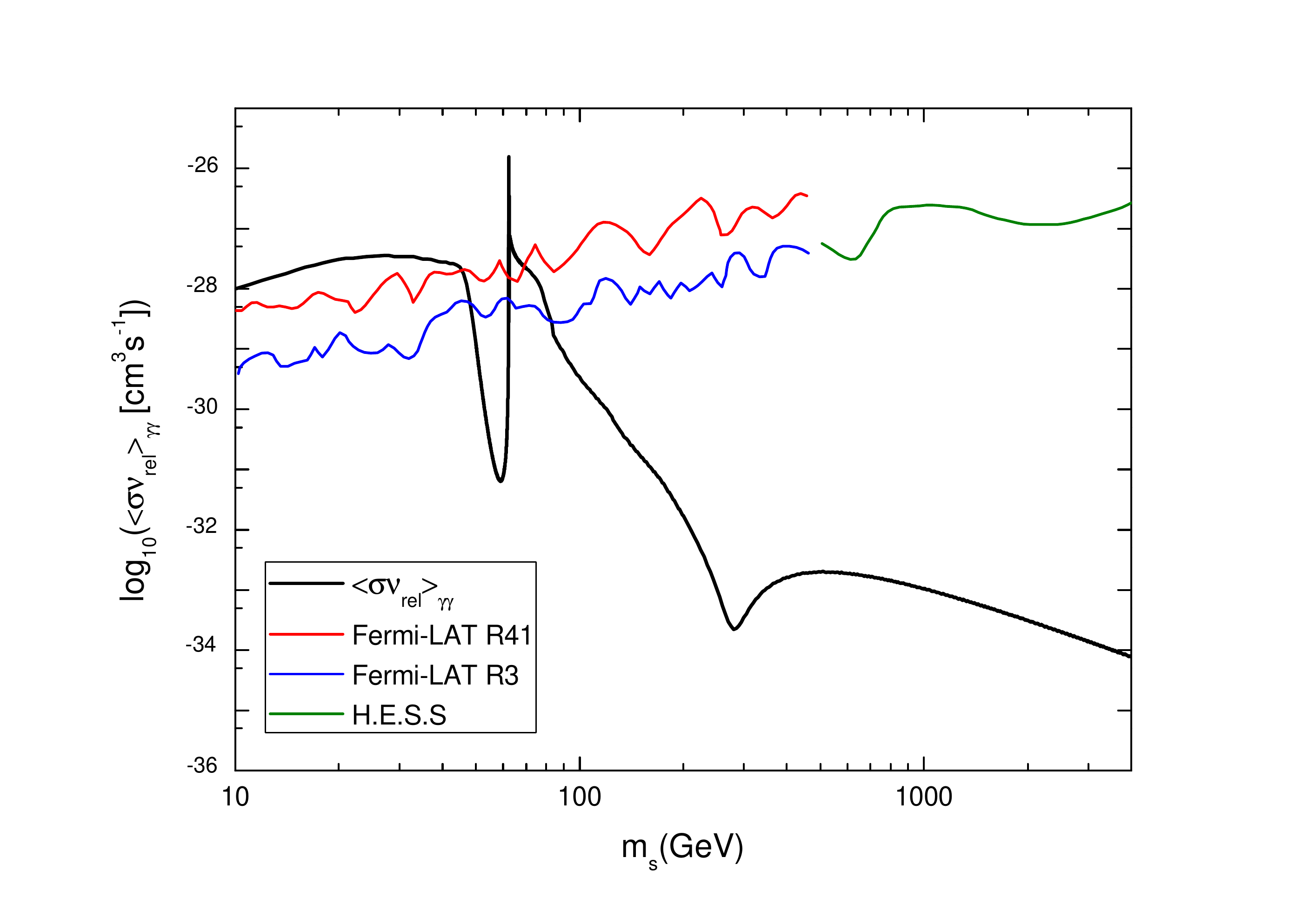}
\vspace{-0.3cm}
 \caption{Constraint from gamma ray spectrum at Fermi-LAT and HESS in the model A ($\mathbf{left}$) and model B ($\mathbf{right}$), respectively.}
 \label{gammaray}
\end{figure}

The value of $<\sigma_{\gamma\gamma} \upsilon_{\text{rel}}>$ is calculated via the standard formula \cite{Average},
\begin{eqnarray}{\label{average}}
<\sigma_{\gamma\gamma} \upsilon_{\text{rel}}>=\frac{x}{16m_{s}^{5}K^{2}_{2}(x)}\int^{\infty}_{4m^{2}_{s}} ds \sqrt{s-4m^{2}_{s}}sK_{1}\left(\frac{x\sqrt{s}}{m_{s}}\right)\sigma_{\gamma\gamma} \upsilon_{\text{rel}}
\end{eqnarray}
where $x=m_{s}/T$, $s$ is the square of the center-of-mass energy, 
and $K1$ and $K2$ are modified Bessel functions of the second kind.
For the details on the calculation of $\sigma_{\gamma\gamma} \upsilon_{\text{rel}}$ in Eq.(\ref{average}),
see the appendixes in Refs. \cite{1412.1105, 1509.04282}.
Similar to the DM-nucleon scattering cross section the value for $\sigma_{\gamma\gamma} \upsilon_{\text{rel}}$ in the model B will be doubled in comparison with A,
which implies that the gamma ray limits will be more sensitive to this model.

In fig.\ref{gammaray} we present the gamma ray constraint on the DM mass.
The Fermi-LAT $R3$ and $R41$ limits are both shown simultaneously.
For the model A, the Fermi-LAT $R3$ limit excludes DM mass above $62.3$ GeV in the resonant mass region,
while the HESS limit is not so sufficient as the LUX limit in the large mass region.
This result is consistent with the earlier one obtained in \cite{1509.04282}.
For the model B, the Fermi-LAT $R3$ limit excludes DM mass above $62.2$ GeV in the resonant mass region,
and the large mass region is not sensitive to the HESS limit similar to the model A.
Note that the difference between the Fermi-LAT $R3$ and $R41$ limits for the DM mass bounds in the resonant mass region is only about $\sim 0.1$ GeV.
In summary, for individual model the nowadays gamma ray and LUX limits set the DM mass upper and lower bound respectively
in the resonant mass region;
and the LUX limit sets the lower mass bound in the large mass region.
The future Xenon1T result will shed light on both these two regions.

\begin{table}[htbp]
\begin{center}
\begin{tabular}{ccccc} 
\text{Model} & $\text{LUX 2016}$ & \text{LHC}~~ & \text{Fermi-LAT}~~ & \text{Mass~Region}\\
\hline\hline
\text{A} & $54.9-62.8$ ~~~~  &  $\geq 52.3$ ~~~~&  $\leq 62.3$ &  $54.9-62.3$ \\
 &  $\geq 834$ & $-$ & $-$ & $\geq 834 $  \\
\text{B} &  $53.8-62.8$~~~~ & $\geq 51.5$ ~~~~&  $\leq 62.2$ & $53.8-62.2$ \\
 &  $\geq $ 3473 & $-$ & $-$ & $\geq 3473$\\
\hline
\end{tabular}
\caption{The DM mass bounds for each experiment. The resonant mass and large mass regions in the individual model are summarized.}
\label{summary}
\end{center}
\end{table}

\section{Conclusion}
In this paper both a real and complex scalar dark matter via the Higgs portal are revisited by the combination of 
the latest direct detections at the LUX experiments and indirect detections at the LHC , Fermi-LAT and HESS experiments.
The DM mass bounds in each experiment  for the two models are summarized in Table.\ref{summary}.
The resonant mass region is further restricted to a narrow window between $54.9-62.3$ GeV for the model A,
which is similar to the model B.
Meanwhile, the large mass region is excluded until $834$ GeV and $3473$ GeV for the model A and B, respectively.
In comparison with earlier individual analysis either on the model A or B in the literature, 
our discussions on them are complete from the viewpoint of nowadays experimental status.

\begin{acknowledgments}
We would like to thank H. Han for discussion and the referee for useful comments.
S. Z thanks the KITPC at Chinese Academy of Sciences for its hospitality,
where this work was initiated.
This work is supported in part by the National Natural Science Foundation of China under Grant No.11405015.
\end{acknowledgments}

\end{document}